\documentclass[pra,twocolumn,aps]{revtex4}
\usepackage{graphicx}
\usepackage{tabularx}
\usepackage{amssymb}
\usepackage{amsmath}
\usepackage[caption=false]{subfig}
\newcommand{\bc}{\begin{center}}
\newcommand{\bef}{\begin{figure}}
\newcommand{\eef}{\end{figure}}
\newcommand{\ec}{\end{center}}

\newcommand{\be}{\begin{eqnarray}}
\newcommand{\bi}{\begin{itemize}}
\newcommand{\ei}{\end{itemize}}
\newcommand{\ben}{\begin{enumerate}}

\newcommand{\een}{\end{enumerate}}
\newcommand{\ee}{\end{eqnarray}}

\begin{document}

\newcommand{\bra}[1]{\big\langle #1 |}
\newcommand{\ket}[1]{| #1 \big\rangle}
\newcommand{\braket}[2]{\big\langle #1 | #2 \big\rangle}
\newcommand{\vmed}[3]{\big\langle #1 | #2 |#3 \big\rangle}

\title{Tunneling induced electron transfer between separated protons}

\author{Patricia Vindel-Zandbergen}
\address{Laboratoire Colisions, Agr\'egats et Reactivit\'e, UMR 5589, IRSAMC,  Universit\'e Paul Sabatier, 31062 Toulouse, France}

\author{Christoph Meier}
\address{Laboratoire Colisions, Agr\'egats et Reactivit\'e, UMR 5589, IRSAMC,  Universit\'e Paul Sabatier, 31062 Toulouse, France}

\author{Ignacio R. Sola}
\address{Departamento de Qu\'imica F\'isica, Universidad Complutense, 28040 Madrid, Spain}
\email{isola@quim.ucm.es}

\begin{abstract}
\noindent
We study electron transfer between two separated nuclei using local control theory. By conditioning the algorithm in a symmetric system formed by two protons, one can favored slow transfer processes, where tunneling is the main mechanism, achieving transfer efficiencies close to unity assuming fixed nuclei. The solution can be parametrized using sequences of pump and dump pi pulses, where the pump pulse is used to excite the electron to a highly excited state where the time for tunneling to the target nuclei is on the order of femtoseconds. The time delay must be chosen to allow for full population transfer via tunneling, and the dump pulse is chosen to remove energy from the state to avoid tunneling back to the original proton. Finally, we study the effect of the nuclear kinetic energy on the transfer efficiency. Even in the absence of relative motion between the protons, the spreading of the nuclear wave function is enough to reduce the yield of electronic transfer to less than one half.
\end{abstract}
\maketitle

\newpage

\section{Introduction}








Quantum control 
can be described as a dynamic process
that prepares coherent superpositions of Hamiltonian eigenstates manipulating the 
amplitudes and relative phases \cite{Shapiro_Brumer_2012,Rice_Zhao_2000,Brif_Rabitz_2011}. 
When this wave function involves several electronic states, the phases
typically oscillate in the sub-femtosecond (hence attosecond) time scale.
However, this is not always the case, as energy differences between Rydberg
states vary on the order of $\sim 1/n$ Hartrees, which can be relatively 
small for large $n$. Hence the period of motion associated to these phase 
oscillations can be of the order of femtoseconds or larger \cite{PhysRevLett.64.2007,PhysRevLett.83.4963}.  

In molecules, other scenarios 
of electronic changes associated to the femtosecond scale exists
whenever the electronic states become quasi-degenerate. This is the
case in the proximity of conical intersections, \cite{corrales2014control,sussman2006dynamic,PhysRevLett.113.113003}
and it is also the
case in the dissociation limit, where many molecular electronic states 
correlate with the same atomic (or fragment) electronic states.
The latter situation is particularly interesting in symmetric arrangements.
Then the initial and target wave function may only differ by the
phase of the initial superposition, which identifies different isomers
that can be converted via tunneling \cite{Sunderman_JPCA1998,Schreiner1300}. 
However, the similar time scales of vibrational and electronic motion
can make the control of the electronic processes particularly
sensitive to the nuclear displacements or even require fully
correlated electron-nuclear motion \cite{Sola_Palacios_Chang_2015,Sola_JCP2013}.
Electronic processes in the attosecond regime may be more protected
against vibrational motion, particularly if the process occurs a
single time \cite{VindelZandbergen2016}. 
However, the effect of vibrational decoherence must still be carefully
studied \cite{PhysRevLett.115.143002,calegari2014ultrafast}. 
%

In this work we investigate electronic transfer between two separated
protons, where the electron is initially in a single proton, breaking
the symmetry of the system.
We have recently shown that a local control (LC) approach \cite{tannor_laser_1999,Rev_LCT_2009,malinovsky1997}
can be used
to find ultrashort pulses that induce electron transfer in very few femtoseconds, yielding
pulses characterized by a prominent (very intense) spike that maximizes 
the probability of retrapping the electron at the desired proton, 
after moving and spreading in the ionizing continuum \cite{VindelZandbergen2016}.
In principle there are infinite solutions of the control problem,
and the LC method is flexible enough to find different types of
solutions.
In this work we show that for particular choices of observables, 
varying the initial conditions can lead to optical control of
electron transfer that explore a different control mechanism,
characterized by slow electron transfer via tunneling.

The paper is organized as follows. In section II we introduce the model 
Hamiltonian and describe the numerical methods used to simulate and control
the dynamics.
In section III we find the control mechanism that implies slow electron 
transfer between two protons largely separated 
via tunneling. 
In section IV we study the role of the nuclear motion in the control of the electron transfer in this timescale. 
Finally, section V is the conclusions. 

\section{Numerical Methods}

We need to use a consistent model for treating both continuum
and bound electronic states in a system with a single electron and two protons.
As a first approximation, we use a $1+1$D Hamiltonian, including
the internuclear distance $R$ and the electron separation to the center
of mass $z$, where the electron is constrained to move in the molecular axis.
For this reduced dimensional study the inter-particle interaction is
modeled by a soft-core Coulomb
potential \cite{javanainen_numerical_1988}.
In the presence of a linearly polarized external field, ${\cal E}(t)$,
and neglecting small mass polarization terms,
the Hamiltonian in the length gauge is 
(atomic units are used throughout unless otherwise stated)

\begin{equation}
\label{H_eq}
H = -\frac{1}{2}\frac{\partial^2}{\partial z^2} 
-\frac{1}{M}\frac{\partial^2}{\partial R^2} + V(z,R) + z{\cal E}(t) 
\end{equation}

where $M$ is the mass of the proton, with the soft-core Coulomb potential 

\begin{equation}
\label{V_eq}
{V(z,R)}=-\frac{1}{\sqrt{1 + (z-R/2)^{2}}} - \frac{1}{{\sqrt{1 +
(z+R/2)^2}}} + \frac{1}{R}
\end{equation}

This model has been extensively applied as a first qualitative step to 
analyze ionization processes in H$_2^+$ and high-harmonic spectra
\cite{PhysRevA.44.5997,PhysRevA.53.2562}, as well as electron-nuclear
dynamics\cite{Sola_Palacios_Chang_2015,Sola_JCP2013,PhysRevA.84.021401,PhysRevA.89.023403}.

Initially, we assume a fixed nuclei approximation, where an hydrogen atom 
and a proton are largely separated. In Section III we select results
for fixed internuclear distances of $10$ and $20$ a.u.
We achieve electron transfer applying LCT. 
The objective is mathematically expressed as the population 
in a target state $|\psi_f\rangle$, constructed as 
a wave function localized at the proton where we want the electron to be
recaptured\cite{Grafe_PRA2005,Kritzer_CPL2009}.
Therefore, the control field depends on the projection on a target state 
\begin{equation}
\label{Target}
{\cal E}(t)=\lambda \Im \!\left[
\vmed{\Psi(t)}{\mu}{\psi_f}\braket{\psi_f}{\Psi(t)}\right]
\end{equation}
where $\Im$ stands for the imaginary part and $\Psi(t)$ is the wave function
of the system. Here $\lambda$ enters as a free parameter to be found numerically,
that characterizes the strength of the laser interaction. 

In finding the local control field with Eq.(\ref{Target}), the projection 
operator $P_t = \ket{\psi_{R_1}}\bra{\psi_{R_1}}$ must commute with the 
Hamiltonian of the system\cite{Rev_LCT_2009}. 
Therefore, the target wave function must be an eigenfunction of the 
Hamiltonian. However, if the
separation of the protons is not large enough, the localized wave functions
are not true eigenstates, as the tunneling time cannot be neglected.
One way of solving this problem is to add a 
very small static field component, ${\cal E}_{DC}$, 
that breaks the symmetry of the Hamiltonian, such that the effective potential 
is tilted, $V_\mathrm{tilted}(z)=V(z)+z{\cal E}_{DC}$. 
Then the target and the initial states are the ground or
first excited electronic wave functions of the Hamiltonian with the DC
component localized at the desired proton. 
The initial state, $\psi_{L}$, is localized at the left potential well and 
$\psi_{R}$ is the target state, localized at the right potential well.

By making ${\cal E}_{DC}$ small enough, the tilted potential has no 
significant impact on the search of the local control field for large
internuclear distances ($R \ge 20$ a.u.). However, the DC component is
an essential ingredient in the control of electron localization at smaller
proton separations.

In our simulations, the initial state is created by exciting the
H$_2^+$ molecule, that is, $\psi_{L}$ is multiplied by an exponential factor
that gives an initial momemtum in the positive direction 
\begin{equation}
\label{mom_eq}
 \Psi(z,0)=\psi_{L_1}(z) \mathrm{e}^{ik_ez}
\end{equation}

In addition, to initiate the LCT approach one needs a small ''seed'' of
population in the right potential well (the target state), 
which we fix as $\approx 0.3\%$. Once the local control field is found this
''seeded'' population is no longer needed, and the simulations shown in the 
results imply $100$\% population in the ground (localized) state at initial 
time.

Numerical results are obtained by solving the TDSE with the 
Split-Operator method \cite{feit_solution_1982,feit_solution_1983,feit_wave_1984} with time steps ranging from
$\Delta t=0.1$ to $0.01$ a.u. depending on the simulation.
A grid of $1024$ points spanning from $z=-80$ to $z=80$ a.u. is used
for the electronic coordinate. Imaginary (''optical'') potentials
\cite{macias_optimization_1994, palao_simple_1998} absorb the outgoing
wave functions avoiding reflection on the grid boundaries and allowing
to measure the ionization probability.
The eigenstates $\psi_{L,R}$ are computed 
using the Fourier Grid Hamiltonian method \cite{marston_fourier_1989}.
The dynamical mechanism of the transfer is studied by analyzing the
approximate phase-space representation of the wave functions at different
times, using the Husimi transformation \cite{Husimi}.

Finally, to study the role of the nuclear motion in the control of the
electron transfer, $1+1$D calculations were performed using the full
Hamiltonian of Eq.(\ref{H_eq}). 
The initial wave packet is the product of the electronic wave function times 
a nuclear Gaussian wave packet $\psi_\mathrm{nuc}(R)$, centered at the left 
nuclei. 
In these simulations we use a grid of $1024$ points ranging from $R = 0.1$ to $R = 150$ a.u. 
for the nuclear coordinate and $256$ points, from $z = -80$ to $z= 80$ a.u., 
for the electronic coordinate. 

\section{Slow electron transfer}

\subsection*{The tunneling mechanism}

We first study electron transfer when the electron stars with a small positive
average momentum. The initial wave function, localized at the left potential,
has an initial momentum of $k_e=0.001$ a.u. [see Eq.(\ref{mom_eq})].
Under these conditions, best results are obtained for nuclei separated
less than $20$ a.u.  
The results of a typical LCT calculation are shown in Fig.1(a) and (b) for
fixed nuclei separated $R=10$ and $R = 20$ a.u. In the latter case,
a free electron transfer (without any acting force) 
would take roughly $t = R/ k_e = 2\cdot 10^4$ a.u. $\approx 500$ fs. 
The initial kinetic energy is not enough to overcome the net attractive force of the 
potential (that is, to overcome the Coulomb barrier between the protons), 
so that in principle the pulse must act to excite the electron and then
to retrap it.

We have used an LCT approach based on the projection operator on 
a target wave function that is the lowest energy eigenfunction localized on
the right well of the tilted potential.
The calculations were performed using a static field of ${\cal E}_{DC}=-5\cdot
10^{-3}$ a.u. and ${\cal E}_{DC}=-2\cdot
10^{-5}$ a.u., for $R=10$ and $R=20$ a.u., respectively.  
Best results were obtained with $\lambda=0.2$ for $R=10$ a.u. 
and $\lambda=2.8$ for $R=20$ a.u. in Eq.(\ref{Target}).

\begin{figure}
\begin{center}
\includegraphics[width=9.0cm]{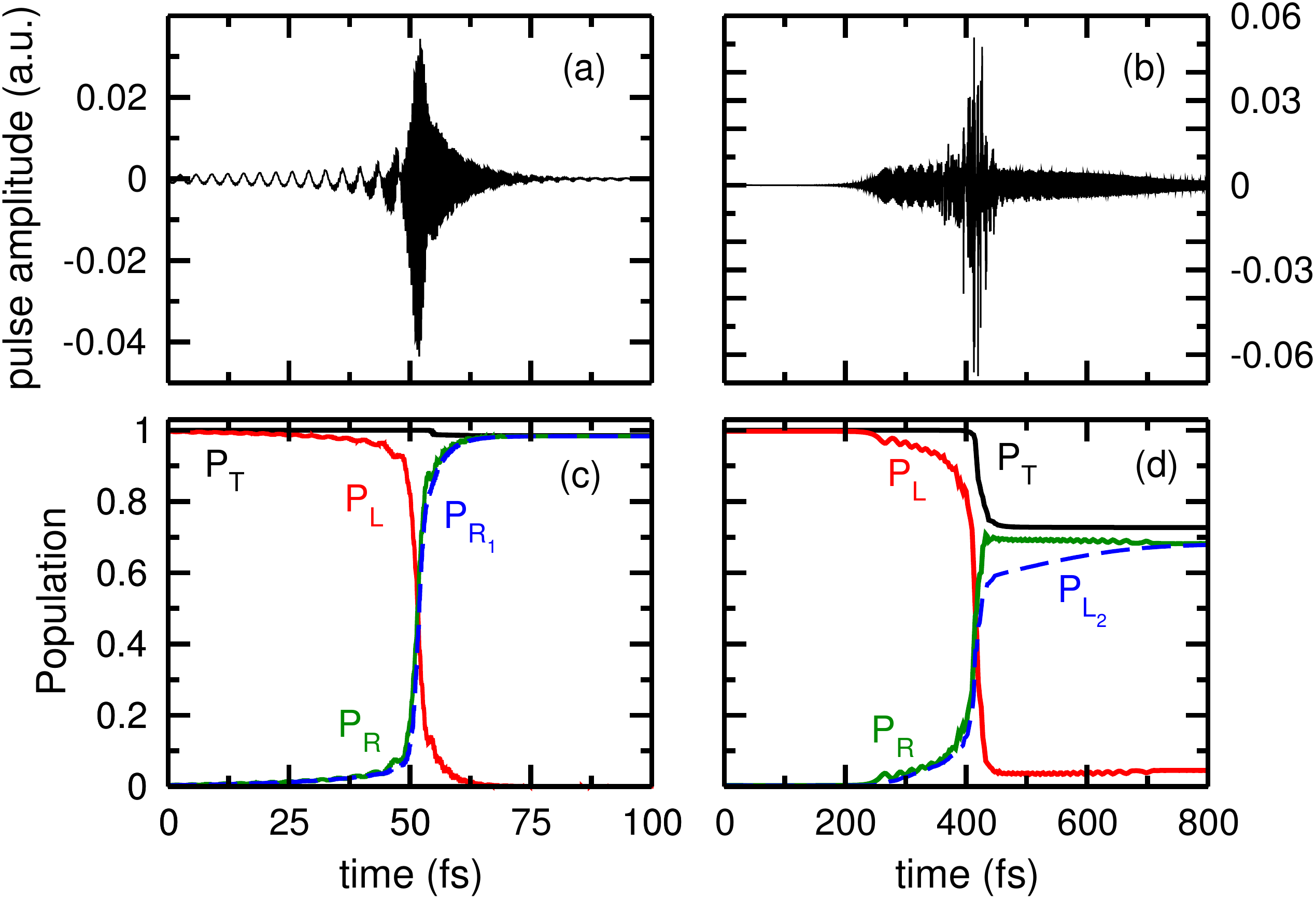}
\caption{Laser control fields and population dynamics for an internuclear distance of $10$ a.u. 
[(a) and (c)] and $20$ a.u. [(b) and (d)], when tunneling is the main mechanism responsible 
for electron transfer.}
\label{pop_laser_slow_R_10_20}
\end{center}
\end{figure}

Fig.\ref{pop_laser_slow_R_10_20}(a) and (b) show the optimal fields for $R=10$ a.u. 
and $R=20$ a.u., while the (c) and (d) panels represent the respective population dynamics 
partitioned into left and right domains

\begin{equation}
\label{pop_eq}
 P_{D}(t)= |\braket{\Psi(z,t)}{\Psi(z,t)}_{z\in D}|^2    
\end{equation}

where $D$ is $(-L/2,0)$ or $(0,L/2)$ (with $L$ the grid size) for the
left and right domains, respectively. Also shown is the yield of the process, 
measured as the overlap of the wave function with the target state,
\begin{equation}
\label{proj_eq}
 P_{R_1}(t)=|\braket{\psi_{R_1}}{\Psi(z,t)}|^2
\end{equation}

As observed, for $R=10$ a.u., full electron transfer is achieved with 
$\sim0.08\%$ population remaining in the left potential well. The 
final population in the right potential well is completely localized in the 
target state, $P_{R_1}/P_R = 1$, an there is no population loss due to ionization.
For $R=20$ a.u., again the electron transfer is almost perfect, with
$\sim6\%$ population in the left hydrogen. The
final population is localized in the
target state, as $P_{R_1}/P_R = 0.99$, while the remaining population (less
than $30$\%) is lost as ionization.

To interpret the mechanism under the electron transfer process, it is
important to notice that the average energy of the electron never
exceeds the energy of the Coulomb barrier between the two protons
in the soft-core Coulomb potential. 
In addition, at these internuclear distances, the tunneling times between
localized states 
are within the time-scale
of the control process. A rough calculation for $R=20$ a.u. gives $t_1 \approx3$ ps 
for population inversion between the ground localized states, $\psi_{L_1}$
and $\psi_{R_1}$, and $t_2 \approx 250$ fs for population inversion between the 
first excited localized states in each well, $\psi_{L_2}$ and $\psi_{R_2}$.
With $R=10$ a.u., the population inversion between $\psi_{L_1}$
and $\psi_{R_1}$ is $t_1 \approx100$ fs.
For the soft-core model other excited states have energies above the Coulomb
barrier.
Roughly, we propose the following mechanism as the key process governing 
the electron transfer controlled by the LC pulse:
First, as a net positive momentum is given to the electron initially,
the electron finds itself distributed between the excited states of the
left hydrogen with energies below the continuum.
The electron is then transferred to the right proton by tunneling. Finally,
 the pulse takes energy away from the electron sitting in the right
proton, effectively stopping the back-tunneling process to the left proton.

\begin{figure}
\begin{center}
\includegraphics[width=9cm]{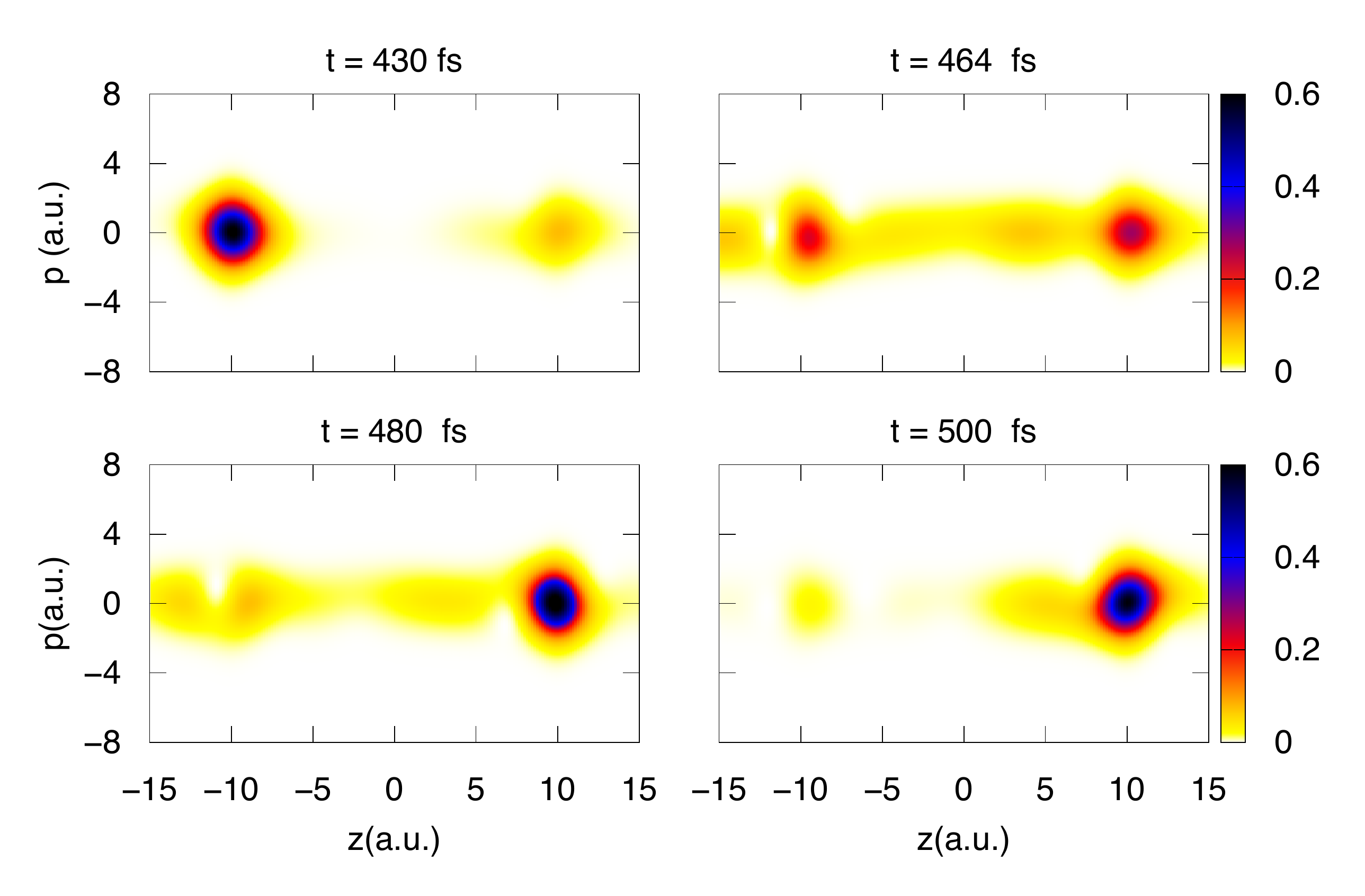}
\caption{Husimi distributions at different times of the electron transferred from the left to the
right proton, for a fixed internuclear distance of $R=20$ a.u.}
\label{Husimi_R20_tunnel}
\end{center}
\end{figure}
 
To help visualizing the process we calculate the Husimi distributions of
the wave functions at different times.
Fig.\ref{Husimi_R20_tunnel} shows the Husimi plots of the electron wave
function at the time the electron transfer is happening for $R=20$ a.u.
As observed, the momentum distribution is localized around zero and does 
not change, implying that the electron is not reaching the continuum while it
 moves to the right potential well.
These distributions correspond to a tunneling mechanism, where the electron 
density starts to ``disappear'' from the left hydrogen atom and 
``appears'' at the proton on the right. 

\begin{figure}
\begin{center}
\includegraphics[width=8cm]{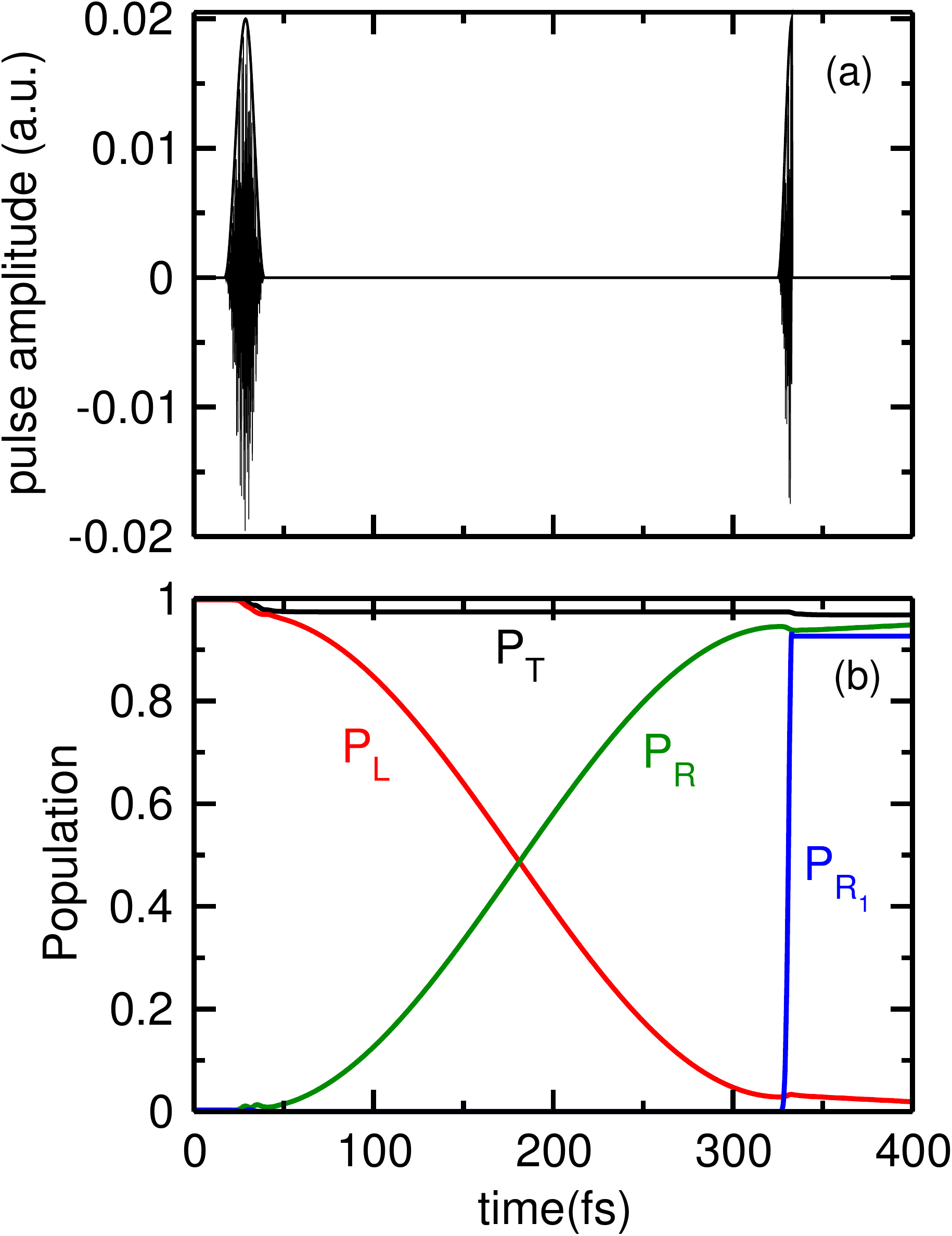}
\caption{Population dynamics and laser pulses used in the "four-state" control scheme. 
$P_{L/R}$ refers to the probability of finding the electron at the left/right 
of the center of mass of the system, $P_T$ is the norm of the wave function (deviations
from unity are due to ionization) and $P_{R_1}$ is the population in the target state.}
\label{poplaser_4states_analytic}
\end{center}
\end{figure}

\begin{figure}
\begin{center}
\includegraphics[angle=-90,width=8.5cm]{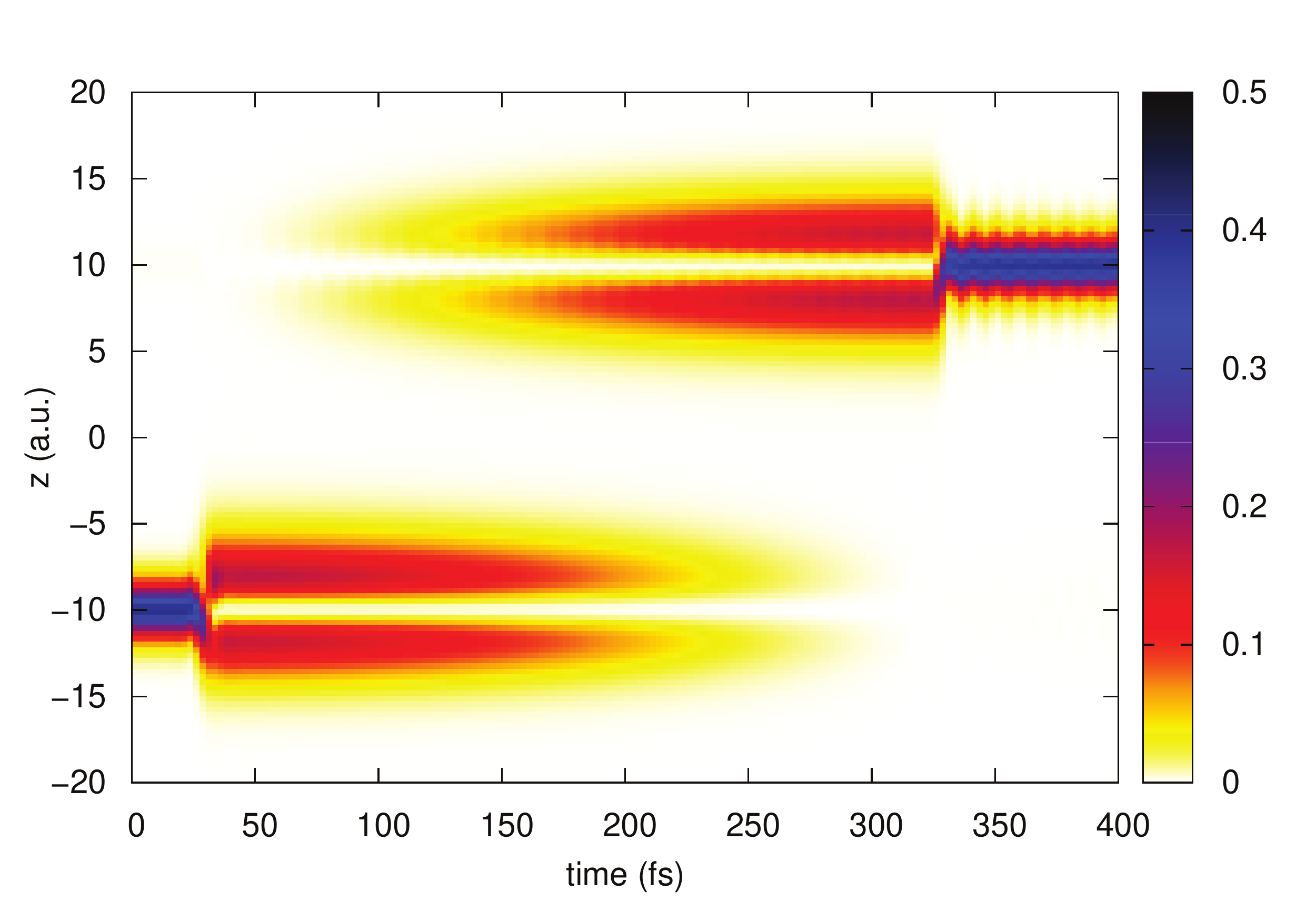}
\caption{Dynamics of the electron density using the "four-state" control scheme.}
\label{wf_tunnel_4states}
\end{center}
\end{figure}

Since tunneling is the main mechanism behind the control process, in the
following we impose such mechanism by choosing sine squared 
($\sin^2(\pi(t-t_0)/\tau)$) shaped pulses that
lead the different steps proposed above for the electron transfer. The
peak amplitude and pulse durations ($\tau$) are estimated numerically by
trial and error to maximize the overall yield. 
%
The idea is to use two laser pulses tuned to the first electronic transition, 
time-delayed by $t_2$, the time it takes for tunneling in the first excited localized
states. 
This is a ''four-state'' scheme where only the localized ground states
($\psi_{L_1}$ and $\psi_{R_1}$) and first excited states ($\psi_{L_2}$ 
and $\psi_{R_2}$)
participate in the dynamics.
In this control scenario we have used $k_e = 0$. 
Fig.\ref{poplaser_4states_analytic}(a) and (b) show the laser pulses and the population dynamics.
The first laser pulse that produces population inversion from $\psi_{L_1}$ to
$\psi_{L_2}$ is equivalent to a $\pi$ pulse of $\sim20$ fs duration,
$0.02$ a.u. pulse amplitude and $\omega=0.39$ a.u.
The second laser pulse that de-excites the electron from $\psi_{R_2}$ 
to $\psi_{R_1}$ is also a $\pi$ pulse of $\sim10$ fs duration, $0.02$ a.u. pulse amplitude and $\omega=0.39$ a.u. 
The yield of population inversion for the second pulse is only slightly lower
than $1$.
Using this scheme, one can completely avoid ionization ($\approx2\%$). 
In fig.\ref{wf_tunnel_4states} we observe the propagation of the electronic wave packet, 
that shows the mechanism of the "four-states" scheme.

For $R=10$ a.u. the mechanism is even simpler. Since the tunneling time in the
ground state is $t_1=100$ fs, starting from the localized state at the left hydrogen, 
we just need to wait enough time for the electron to completely transfer to the right 
proton, without the action of an external laser field. However, in this case, the
opposite process also occurs, so that the electron cannot remain localized.

In summary, electron transfer between protons that are separated by 
moderately large distances
is possible by means of a slow transfer process, where tunneling below
the internal barrier is the predominant mechanism.
Tunneling can be made the sole mechanism responsible for the process.
In this case, however, local control theory is not required, 
as analytic pulses, or even no external field, can lead to the same final results.
The tunneling mechanism can not be used effectively when the internuclear
distance increases, as the tunneling time increases exponentially and
one needs to selectively excite the electron to Rydberg states ever closer
to the continuum, a process that is difficult by itself. 
In addition, as we show in the next section, the nuclear dynamics acts as a 
very strong perturbation source that affects the yield of the process.

\subsection*{The role of the nuclear motion}

\begin{figure}
\begin{center}
\includegraphics[width=8cm]{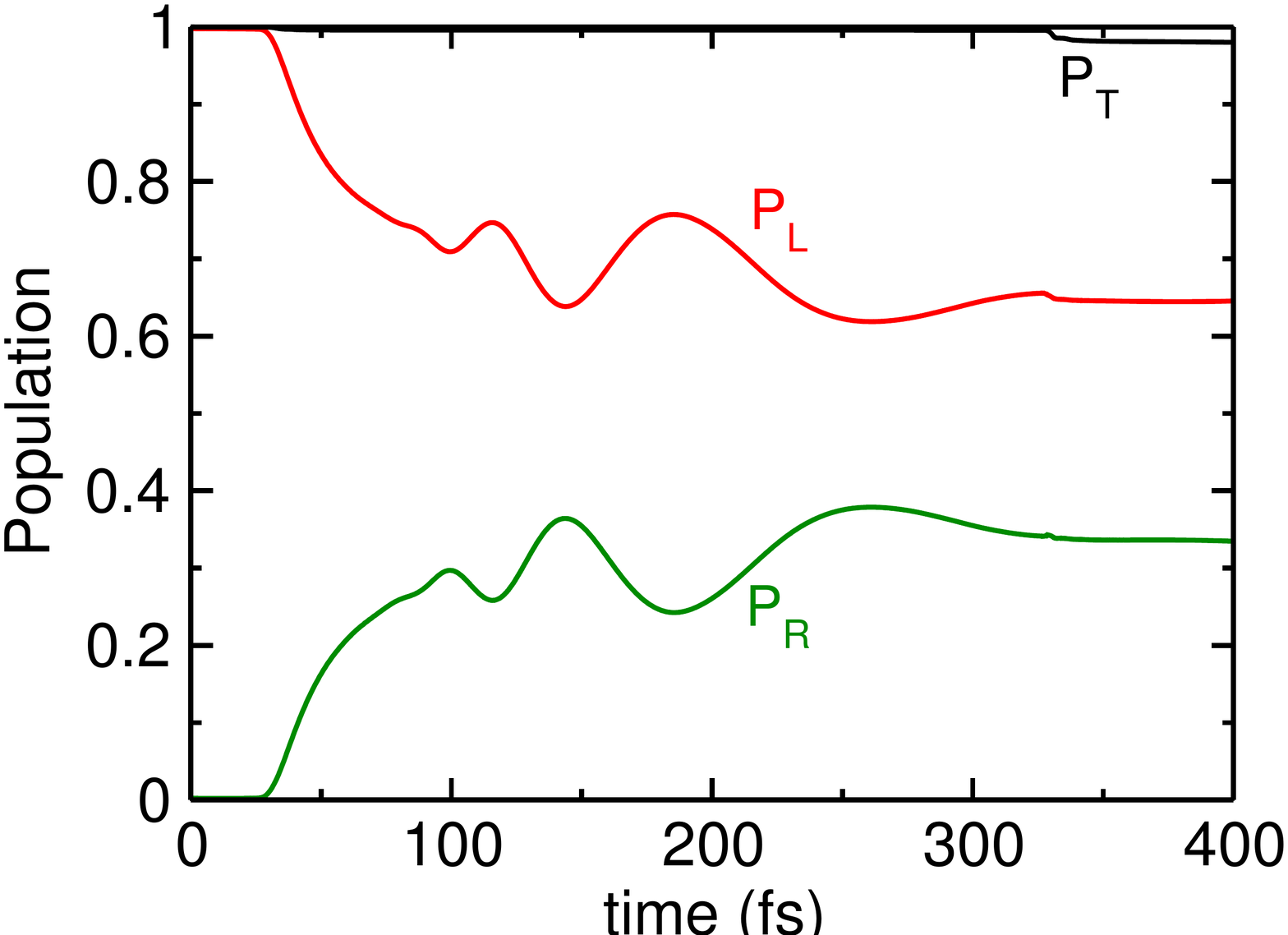}
\caption{Population dynamics in a 2-D calculation with zero average initial
momentum using the "four-state" scenario.}
\label{pop2D_R20}
\end{center}
\end{figure}

\begin{figure}
\begin{center}
\includegraphics[width=7cm]{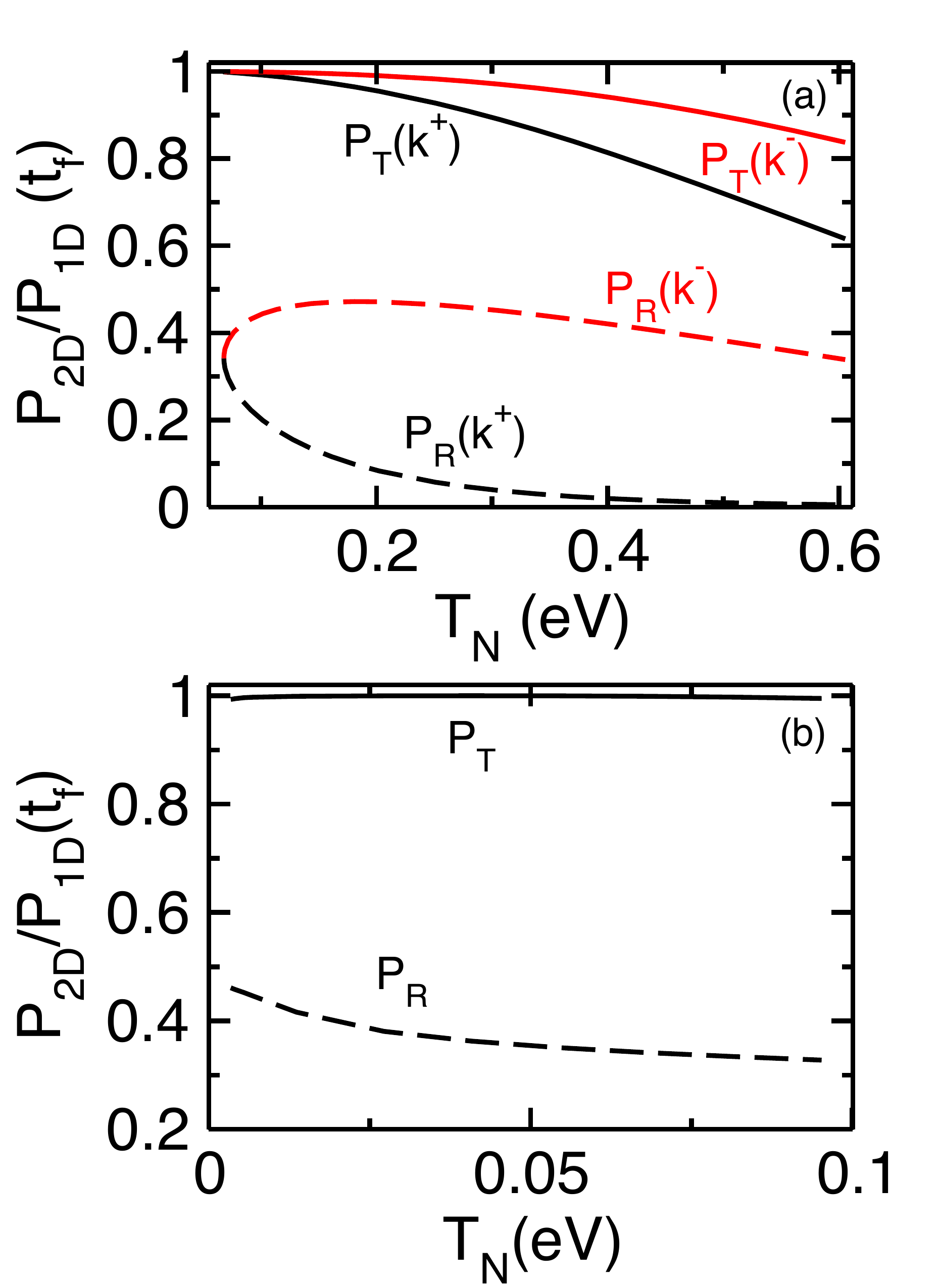}
\caption{Variation of the probability of finding the electron at the right proton at final time $P_R$ and
the norm of the wave function $P_T$ as a function of the initial nuclear kinetic energy 
(a) depending on the initial average momentum given to the nuclei (positive, $k_{n}^+$, 
or negative, $k_{n}^-$) and (b) depending on the initial width of the nuclear wave function,
assuming zero average initial momentum}
\label{Tcin_vs_yield}
\end{center}
\end{figure}

The fixed nuclei approximation is valid as long as the electronic processes
occur in a time-scale much faster than that of the nuclear motion.
In the slow electron transfer mechanism, the transfer times are of the
order or larger than typical vibrational periods.
Here, we analyze how the nuclear motion affects the yield of the tunneling process.
We use the optimal pulses found in the fixed-nuclei LCT approach, as well as the 
analytic pulses of the "four-state" scheme and 
apply them on a full $2$-D (or ($1+1$D)) calculation considering different initial 
nuclear wave functions. The effect of the initial nuclear kinetic energy
is considered in two different ways: by studying the effect of the width 
of the initial Gaussian nuclear wave function and by adding a net momentum in
the positive or negative direction.
However, we do not directly apply the LCT approach to the ($1+1$)D TDSE,
that is, the pulses are not directly optimized taking into account the nuclear
motion.

\begin{figure}
\begin{center}
\includegraphics[width=9cm]{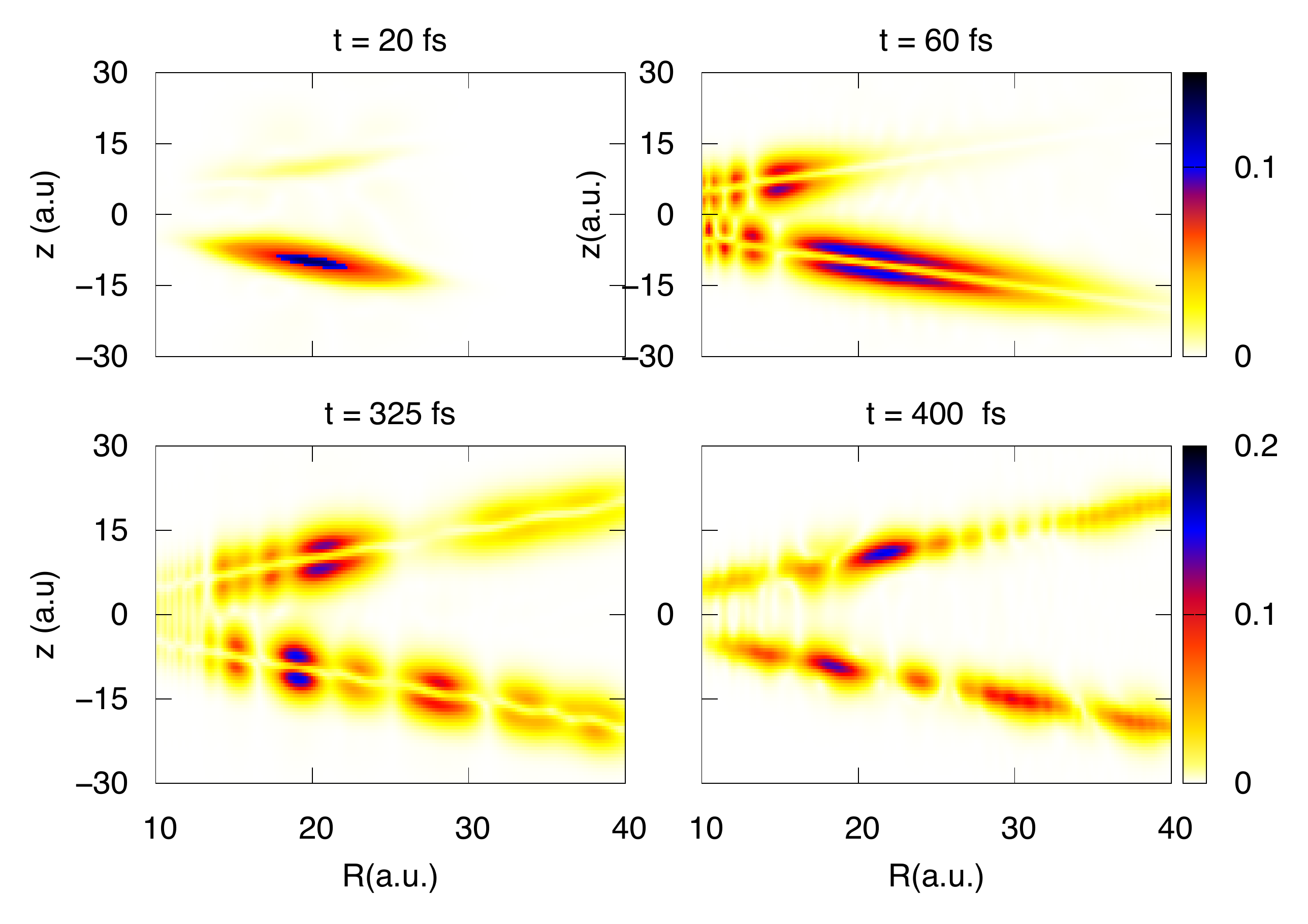}
\caption{Snapshots of the electron density at different times. 
The electron transfer corresponds to the "four-state" scheme 
described in the 1D approach, where an initial laser pulse of 
$\sim 20$ fs duration, $0.02$ a.u. amplitude and $\omega=0.39$ a.u., 
is used to excite the electron to the first excited state localized in the left hydrogen. 
A second laser pulse with same amplitude and frequency but $\sim10$ fs 
duration de-excites the electron to the ground state 
localized in the right nucleus. The time-delay between these two $\pi$ pulses corresponds to the tunneling time between the two first localized excited states.
Although the wave packet spreading along the internuclear 
distance is a main issue affecting the photoassociation process, 
some degree of population transfer is achieved and ionization is avoided.} 
\label{wf_2D_kn0}
\end{center}
\end{figure}

As a representative example, fig.\ref{pop2D_R20} shows the effect of the nuclear motion over the population 
dynamics for $R=20$ a.u., when the analytic pulses of the 1-D "four-states" scenario are used 
in a 2-D calculation and no net average momentum is given to the nuclei. 
Although some population transfer is achieved and ionization is also avoided, the effect of the nuclear motion completely influences the photoassociation process.  
Now, only $\sim30\%$ population remains at the right Hydrogen and the electron transfer mainly occurs in the first $150$ fs, but the transfer mechanism is the same as in the 1-D situation. 
As observed in fig.\ref{wf_2D_kn0}, the 2-D wave packet initially spreads
along the internuclear coordinate during the first $20$ fs. 
Then, the first laser pulse excites the electron to $\psi_{L_2}$ 
in the left hydrogen (negative $z$ values). 
Because at $R = 20$ a.u. there is a small attractive force between the nuclei,
even when no net momentum is given, the nuclei approach.
Now, the tunneling time is lower since the wave function
is moving to lower internuclear distances (breaking the degeneracy
of the electronic wave functions due to the internal barrier),
and the population transfer mainly finishes at $t \sim 60$ fs.
Between $325$ and $330$ fs the population inversion 
occurs from the excited to the ground state of the right proton,
when the second laser pulse is acting. 
Finally, we obtain an electron population distributed in the two potential wells that spreads 
along the internuclear distance, but mainly around $R=20$ a.u.

Fig.\ref{Tcin_vs_yield} shows the effect of the nuclear motion as the ratio between the
$2$-D and the $1$-D results for the yield of the process $P_{R}(\infty)$ 
(the probability of being localized at the right proton at final time) 
as well as for the remaining (not ionized) population $P_T(\infty)$.
In Fig.\ref{Tcin_vs_yield}(a) we fix the initial width of the nuclear Gaussian wave packet
at $\sigma = 0.31$ a.u. and we consider the effect of positive or
negative nuclear momentum. In fig.\ref{Tcin_vs_yield}(b) the net momentum is zero and
only the width of the initial Gaussian wave packet is changed.
As a first insight, one can observe the yield of population transfer decreases to $\sim30$\% when no initial momentum is given to the nuclei, but the total population remains as in the 1-D calculations.

Also, we can notice differences in the variation of the final population
 depending on the sign of the initial nuclear momemtum. 
The total population decreases as the initial kinetic energy increases
 for both positive and negative momenta, but the degree of population loss is bigger in the positive case.
As well, we can appreciate this behavior in the population transfer to the right potential well for positive values of the nuclear momentum.   
However, when considering the negative case, initially there is an increase 
in the electron transfer to the right proton for small initial kinetic energies of the nuclei. Then, the yield decreases but not as drastically as for positive values.

\begin{figure}
\begin{center}
\includegraphics[width=8cm]{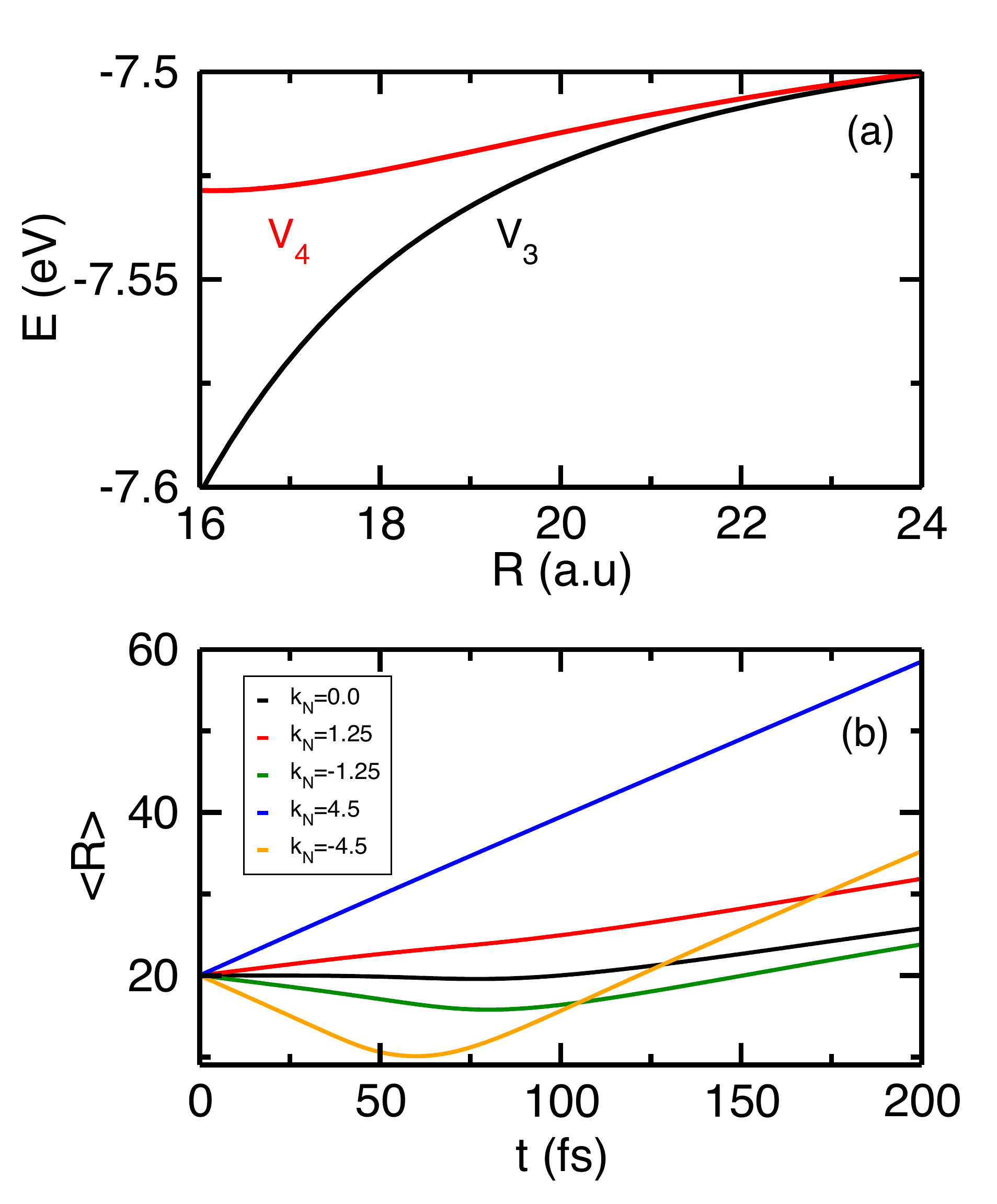}
\caption{(a) Potential energy curves of the second ($V_3$) and third ($V_4$) excited states 
around R=20 a.u. 
(b) Variation of the average internuclear distance for different values of the initial average
nuclear momentum. 
For negative values, the nuclei initially approach until a minimum value of 
$\langle R\rangle$ when they collide after which the internuclear distance 
increases. For positive values of $k_N$, the nuclei always separate, at a
speed that depends on $k_N$. 
(c) Overlap between the nuclear wave functions corresponding to the second, $\psi_3$, 
and third, $\psi_4$, excited states for different initial nuclear kinetic energies.} 
\label{pes_R_S}
\end{center}
\end{figure}

These results can be explained by considering the potential energy curves 
of the second and third excited states where the tunneling is happening 
[fig.\ref{pes_R_S} (a)] and the variation of the average of the internuclear 
distance with the initial nuclear kinetic energy [fig.\ref{pes_R_S} (b)]. 
When we apply a negative momentum, the nuclei start to get closer,
 so the tunneling time decreases, therefore, initially the electron transfer increases.
But, from kinetic values larger than $\sim0.15$ eV the electron transfer 
remains constant, regardless of the increase of the initial nuclear momentum.
The decrease of the population in the right potential is due to the total population loss by ionization.

When considering positive values of the initial momentum, as the nuclei separate, 
the energy difference between the excited states decreases and the tunneling time 
increases, lowering the population transfer, finally even avoiding photoassociation. 
  
Regarding the nuclear potential energy curves around $R=20$ a.u. [fig.\ref{pes_R_S}(a)],
 for lower values of the internuclear distance, 
the energy difference between states increases, so the overlap between 
the second and third excited states must decrease.
 On the other hand, as we move to larger values of R, 
the energy difference becomes nearly zero, so the excited states 
are now degenerated states and the overlap between them is close to $1$.

Resuming, for negative values of the initial nuclear momentum, 
the population transfer increases as the nuclei approximate. 
The internuclear distance decreases so the energy difference between the excited states increases and the tunneling time diminishes.
Positive values of initial kinetic energies, imply a rise of the internuclear distance as the nuclear are separating, so the energy difference between the excited states is reduced. The tunneling time increases so the population transfer is lower. 
This effect is more significant as the initial nuclear kinetic energy is bigger. 
Regarding the effect of the initial width of the nuclear wave packet, 
we can observe the total population remains constant but the population transfer increases as we move to lower values of initial kinetic energy. Therefore, 
highly localized nuclear wave packets (small widths) favor the electron transfer.

\section{Conclusions}

We have studied electron transfer between two separated protons using Local
Control Theory (LCT), in a 1D model system assuming fixed nuclei 
interacting through soft-core Coulomb potentials. 
By properly choosing the control functional, the initial conditions
and the timing of the dynamics, we have searched for optical pulses
that lead to electron transfer on the time-scale of hundreds of
femtoseconds, achieving high yields for the overall process.

The analysis of the transfer showed that the process
was mediated by tunneling below the Coulomb barrier between the
protons. Generalizing the results we have proposed control strategies
where two time-delayed pulses are used first to prepare the initial state 
closer to the barrier whenever the protons are too largely separated
(and hence the tunneling rate is too slow) and finally
to de-excite the electronic state on the target proton, avoiding the
backward transfer. In between the pulses the dynamics follows
by laser-free tunneling, which essentially yields perfect electron
transfer. 

This mechanism is in contrast to recent proposals that used impulsive
ultrashort pulses to mediate the transfer through the ionizing continuumi\cite{VindelZandbergen2016}. 
In the latter case the yields were much smaller, but remained less affected
by the motion of the nuclei. 

Using a 2D study of the three particles
moving collinearly showed that the electron transfer mediated by 
tunneling was severely affected by the vibrational motion of the 
nuclei. Since the tunneling times depend exponentially on the energy
differences, the yield of electron transfer was half that observed 
in the 1D case even for initially frozen nuclei, simply due to the spreading
of the nuclear wave function. The effect of the remaining dimensions
of the systems still needs to be evaluated, although they will likely 
affect in similar ways to the yield of the process. In addition, electron transfer 
through tunneling is more difficult to achieve in non-symmetric systems,
where the ions are different. However, laser-enhanced tunneling is
also a possible solution. Future studies are needed to better assert
the merits of the different electron transfer mechanisms, but presumably
the better choice will depend on the system of study and the initial
kinetic energy of the ions.





\section*{Acknowledgments}
Financial support from the MICINN (Projects CTQ2012-36184 and CTQ2015-65033-P) 
and the COST-action (Grant No. CM1204, XLIC), as well as the computational 
facilities by CALMIP, Toulouse, are gratefully acknowledged.

\bibliography{bibliog_paper}



\end{document}